\documentclass[manuscript]{aastex}
\usepackage{amsmath}
\usepackage{float}

\shorttitle{Regimes of Thermal-Diffusional Pulsation in White Dwarf Flames}
\shortauthors{Guangzheng Xing et al.}

\begin{document}

\title{Thermal-Diffusional Instability in White Dwarf Flames: \\Regimes of Flame Pulsation}

\author{Guangzheng Xing\altaffilmark{1, 2, \dag},
Yibo Zhao\altaffilmark{1, 3, \dag},
Mikhail Modestov\altaffilmark{5, 6},
Cheng Zhou\altaffilmark{1, 4},
Yang Gao\altaffilmark{1, 4, *},
Chung K. Law\altaffilmark{1, 6}}

\affil{$^1$ Center for Combustion Energy, Tsinghua University, Beijing 100084, China}
\affil{$^2$ Department of Electrical Engineering, Tsinghua University, Beijing 100084, China}
\affil{$^3$ Department of Engineering Physics, Tsinghua University, Beijing 100084, China}
\affil{$^4$ Department of Thermal Engineering, Tsinghua University, Beijing 100084, China}
\affil{$^5$ Nordita, KTH Royal Institute of Technology and Stockholm University, 10691, Stockholm, Sweden}
\affil{$^6$ Department of Mechanical and Aerospace Engineering, Princeton University, Princeton , New Jersey 08544, USA}
\affil{$^\dag$ These authors contribute equally to the paper.}
\affil{$^*$ Email: gaoyang-00@mails.tsinghua.edu.cn}

\begin{abstract}

Thermal-diffusional pulsation behaviors in planar as well as outwardly and inwardly propagating white dwarf carbon flames are systematically studied.
In the 1D numerical simulation, the asymptotic degenerate equation of state and simplified one-step reaction rates for nuclear reactions are used
  to study the flame propagation and pulsation in white dwarfs.
The numerical critical Zel'dovich numbers of planar flames at different densities ($\rho=2$, 3 and 4$\times 10^7$~g/cm$^3$)
  and of spherical flames (with curvature $c=$-0.01, 0, 0.01 and 0.05) at a particular density ($\rho=2\times 10^7$~g/cm$^3$) are presented.
Flame front pulsation in different environmental densities and temperatures are obtained to form the regime diagram of pulsation,
  showing that carbon flames pulsate in the typical density of $2\times10^7~{\rm g/cm^3}$ and temperature of $0.6\times10^9~{\rm K}$.
While being stable at higher temperatures,
  at relatively lower temperatures the amplitude of the flame pulsation becomes larger.
In outwardly propagating spherical flames the pulsation instability is enhanced and flames are also easier to quench
  due to pulsation at small radius,
  while the inwardly propagating flames are more stable.

\end{abstract}


\section{Introduction}

The nuclear burning of white dwarf and explosion of type Ia supernova (SN Ia) continues to be a topic of substantial astrophysical interest;
  the solution of which requires the input of knowledge from the fields of combustion, statistical physics and nuclear physics \citep{wheeler2012,oran2015}.
Despite the possibility of merging of double degenerate stars followed by the supernova events,
  the traditional single (sub)Chandrasekhar mass \citep{cha31} C-O white dwarf (WD) burning is still considered to be responsible for a large population of SN Ia \citep{blinnikov2004,hillebrandt2013}.
Specifically, one of the key issues in its modeling is related to the flame acceleration and deflagration-detonation transition (DDT),
  with flame front instabilities being considered as a possible mechanism in driving the acceleration.

In general there are three modes of flame-based instabilities which can fundamentally affect the propagation speed of the global flame front,
  namely \citep{Law03}:
  the Landau- Darrieus instability (L-D instability ) \citep{lan44,bell2004A},
  the Rayleigh-Taylor instability (R-T instability) \citep{taylor1950,Bell44}
  and the thermal-diffusional instability \citep{siva77,byc95,gla13}.
The L-D instability originates from the density discontinuity across the flame front,
  and leads to wrinkled front surface which in turn increases its global propagation speed.
The R-T instability, which also wrinkles the front surface,
  occurs as a result of front acceleration or when it propagates in a gravity field,
  if the density gradient is either opposite to the direction of gravity or in the direction of the front acceleration,
  forming bubbles rising toward the stellar surface in WD while accelerating the flame propagation.
Both of these two instabilities have been extensively investigated as possible triggers of the DDT.
Less studied is the role of the thermal-diffusional instability,
  due to the disparity in the heat and mass diffusivities of the medium,
  in the dynamics of WD nuclear flames,
  and is investigated herein in the context of SN explosion.
While the R-T instability is the dominant hydrodynamic instability governing the flame surface deformation and flame acceleration,
  the thermal-diffusional instability, as discussed below, governs the local flame quenching and could speed up the flame
  in the smaller scale comparable to the flame thickness.
However, since the thermal-diffusional instability can only be captured in simulations in which the reaction zone of the flame,
  which is usually an order of magnitude smaller than the thermal diffusion zone, is well resolved,
  such instability cannot be observed in most of the current 2D simulations and its effects have always been overlooked.
Before going into fine enough grids well resolving the reaction zone thickness,
  it deserve investing the instability regimes through semi-analytical as well as simplified 1D simulations.

The thermal-diffusional instability is based on the strong temperature dependence of the reaction rate
  coupled with the thermal and mass transport properties of the combustible mixture \citep{siva77},
  manifested through the Lewis and Zel'dovich numbers for the stability limits.
While cellular and pulsating instabilities can both be exhibited, depending on whether the controlling Lewis number is smaller or greater than unity,
  the excessively large Lewis number associated with the WD flames, to be demonstrated later,
  implies the dominance of pulsating instability for the present phenomena.
In this regard we note that the pulsating behavior of such instability in WD carbon planar flames was considered by \citet{byc95},
  who suggested that since the average speed of the pulsating flame is lower than that of the stable flame,
  thermal-diffusional pulsation cannot lead to DDT.
However, (1) despite reduced average velocity, the on-pulse flame speed in one period of pulsation can be much larger than the planar flame speed
  \citep{gamezo2014,poludnenko2015}.
In addition, (2) in the off-pulse phase,
  flame speed can be sufficiently reduced so that the flame temperature is low enough for the combustion wave to quench,
  which is relevant for evaluating the condition of spherical flame ignition in WD \citep{williams1985,chen2011}.\footnote{Both computational and experimental results show that the oscillatory extinction has significant difference compared with steady flame extinction \citep{christ2002}. It is also noted that curvature affects the ignition of spherical detonation waves as well \citep{he1994}.}
Furthermore, (3) the curvature of the flames affects the pulsating criterion and the amplitude of pulsations
  \citep[][see also the Appendix B of this paper]{sun02},
  which may be crucial for the burning of the ignition kernels and fuel pockets formed through turbulence or R-T instabilities \citep{bell2004A,zingale2005}.
Finally, (4) the equation of state (EoS) of the combustible medium as well as the reaction rate function modify the critical value
  of the Zel'dovich number.
In this paper we discuss the above issues and elaborate the stability criterion for pulsating WD flames.

\section{Physical conditions for nuclear flames in white dwarf}

In order to describe nuclear flames accurately and investigate their pulsating properties,
  we briefly review the characteristic features of WD matters and nuclear reactions.
Below we discuss the relevant EoS as well as the specific nuclear reaction rate and the corresponding Zel'dovich number.

\subsection{Equations of state and the Lewis number}

The central part of WD exhibits high density, pressure and temperature, and
  the EoS is determined by highly relativistic degenerate electrons \citep{Tim92}.
Following \citet{lan80}, we note the major physical quantities of degenerate Fermi gas assuming ultra-relativistic regimes,
  with the specific internal energy for a relativistic degenerate electron gas determined as:
  \begin{equation}
  E_0=\frac{3}{4\rho}{(3{\pi}^2)}^{\frac{1}{3}}\hbar c_{\rm l}{(N\rho)}^{\frac{4}{3}},
  \label{specific internal energy}
  \end{equation}
  where $N$ is the number of electrons per unit mass, $\rho$ the mass density,
  $\hbar$ the reduced Planck constant and $c_{\rm l}$ the speed of light.


For the degenerate Fermi gas in the simulation, at temperatures much lower than the Fermi temperature
  ($1.6\times 10^{10} \rm K$ at density $\rho=2\times10^7 \rm g/{cm}^3$),
  the specific heat is a linear function of temperature:
  \begin{equation}
  C_{\rm p}=N\frac{(3\pi^2)^{\frac{2}{3}}}{3\hbar c_{\rm l}} \left(\frac{1}{\rho N}\right)^{\frac{1}{3}}k^2T,
  \label{specific heat}
  \end{equation}
  where $k$ is the Boltzmann constant and $T$ the temperature in Kelvin.
Furthermore, $H$ the enthalpy per unit mass is written as $H=E+P/\rho$, where $E=3P/\rho$ in the relativistic regime,
  and $E$ is the specific internal energy at temperature $T$, being
  \begin{eqnarray}
  E=E_0+\int_0^TC_{\rm p} dT=E_0+\frac{1}{2}N\frac{(3\pi^2)^{\frac{2}{3}}}{3\hbar c_{\rm l}} \left(\frac{1}{\rho N}\right)^{\frac{1}{3}}\left(kT\right)^2.
  \label{total energy}
  \end{eqnarray}
Consequently, the enthalpy has the following form:
  \begin{equation}
  H={(3\pi^2)}^{\frac{1}{3}}\frac{\hbar c_{\rm l}}{\rho}{(\rho N)}^{\frac{4}{3}}
  \left[1+\frac{2}{9}(3\pi^2)^{\frac{1}{3}}\frac{\left(kT\right)^2}{(\hbar c_{\rm l})^2}(\rho N)^{-\frac{2}{3}}\right],
  \label{eos}
  \end{equation}
  which will be used later in the analytical and numerical studies of the instability.

Another important parameter in combustion waves is the Lewis number $Le$,
  defined as the ratio of the thermal diffusion coefficient to the mass diffusion coefficient.
The WD thermal diffusion is mainly carried by relativistic degenerate electrons while mass diffusion is attributed to the nearly static nucleons,
  making $Le$ a very large number.
Without going into detailed evaluation, we directly refer to the estimates in \citet{yak80} and \citet{blinnikov2004}, which yield $Le\sim10^7$.
In an environment of such extremely large Lewis numbers, the thermal-diffusional instability is crucial for flames,
  and the analytical results of \citet{siva77} can be applied by changing the EoS from ideal gas to relativistic degenerate gas.

\subsection{Nuclear reaction rate and the Zel'dovich number}

In WD explosions the major energy release comes from the process of nuclear fusion, occurring in the interior of the star.
The most energetic branch corresponds to carbon reaction followed by oxygen reactions.
Due to the extremely high temperatures accompanying fusion, the nuclear reaction rates differ from those of terrestrial chemical reactions,
  and is often described as \citep{fow75}
  \begin{equation}
  \Re=a^2\exp{\left(-\sqrt[3]{\frac{E_{\rm a}}{T_9}}\right)},
  \label{reaction rate}
  \end{equation}
  where $a$ is the molar concentration of the fuel, and ${E_{\rm a}}$ is the activation energy,
  which is $84.165^3$ for pure carbon reaction and $135.93^3$ for oxygen,
  with both the temperature $T_9$ and energy $E_{\rm a}$ in unit of $10^9$~K.
The relatively lower activation energy for the carbon reaction means that carbon will first react when the local temperature is high enough
  for the carbon nuclei to break the energy barrier.
This initial reaction will be followed by the oxygen reaction after the environment temperature is further heated up by the energy released
  in the carbon reaction.
This is also confirmed by the numerical simulations of WD flames \citep[see, e.g.][]{Woo11},
  in which the oxygen flame is found to follow the carbon flame with the same carbon flame speed.
Hence we can study the stability of the carbon flame first, assuming that it represents the leading front of the WD flame propagation.
For simplicity, we assume the WD consists of pure $\rm{{}^{12}C}$ and the simplified one-step reaction
  $\rm{{}^{12}C+{}^{12}C\rightarrow {}^{24}Mg}$ is used instead of the full reaction chain.
Both the simplified reaction rate (\ref{reaction rate}) and the pure carbon reaction (simplified from the carbon-oxygen two stage mechanism) used here are
  to enable the quantitative comparison with existing analytical models for a clear physical insight of the instability.

The temperature sensitivity of a reaction can be measured by the Zel'dovich number,
  which is the key parameter in studying the thermal-diffusional instability.
Following the definition of the Zel'dovich number, \citet{blinnikov2004} showed that
  \begin{equation}
  Ze=\frac{\partial\ln\Re}{\partial\ln T}.
  \end{equation}
By further accounting the relatively moderate temperature increase of the burnt matter with respect of the unburnt mixture, $\theta=T_{\rm b}/T_{\rm u}$,
  and assuming that reaction occurs only in a narrow region within which the temperature deviates from $T_{\rm b}$ on the order of $Ze^{-1}$ \citep{Law03},
  and by adopting the reaction rate (\ref{reaction rate}), the Zel'dovich number has the following form:
  \begin{equation}
  Ze=\frac{1}{3}{\frac{\theta -1}{\theta}}\sqrt[3]{\frac{E_{\rm a}}{T_{\rm b}}}.
  \label{zeldovich}
  \end{equation}
For carbon nuclear flames its typical values are around 10 to 20 (Fig. 1),
  which is in line with the recent simulation of the SN Ia explosion \citep[][]{gla13}.
\\

\begin{figure}[H]
\centering
  \includegraphics[width=0.73\textwidth]{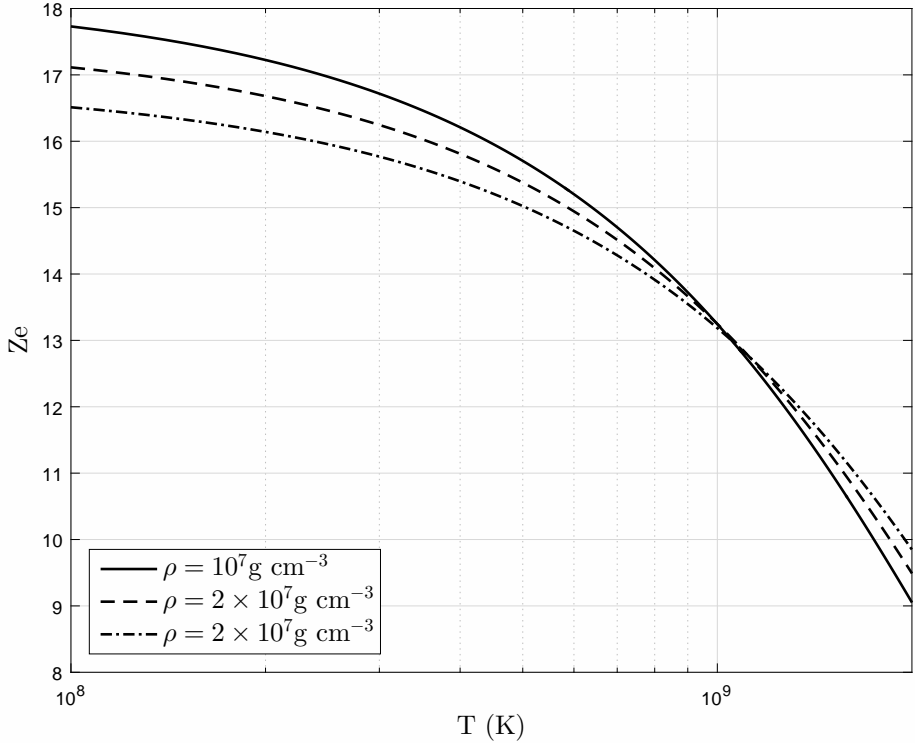}
  \caption{Zel'dovich number for carbon flame as a function of the unburnt temperature for different densities in WD,
  according to Eq.~(\ref{Ze}).}
  \label{fig_ZevsT}
\end{figure}

In the following analysis we assume that the density of nucleon $\rho_{\rm n}$ remains constant during carbon nuclear reaction,
  accordingly the electron density $\rho$ keeps constant.
Noting that while the density does change, at the most ~50\% \citep{Woo11},
  density enters the equations under the cubic root, which further reduces the effect of its variation.
This allows us to obtain the relationship between $T_{\rm b}$ and $T_{\rm u}$ with a certain initial density.
As the enthalpy change equals to the heat release, i.e. $\Delta H= Q_{\rm c}$, we can then calculate $T_{\rm b}$
  according to the EoS (\ref{eos}) as,
  \begin{equation}
  T_{\rm b}=\sqrt{T_{\rm u}^2+\frac{9}{2} {\left(3\pi^2\right)}^{-\frac{2}{3}} Q_{\rm c}  k^{-2}   {\hbar c_{\rm l} }\rho^{\frac{1}{3}} N^{-\frac{2}{3}}},
  \end{equation}
  where the heat release of the carbon reaction is $Q_{\rm c}=5.6\times 10^{17} {\rm erg/g}$ \citep{fow75}.
Then for a given $\rho$ and $T_{\rm u}$ one can calculate the Zel'dovich number (\ref{zeldovich}) as,
  \begin{equation}
  Ze = \frac{28.55\times10^{3}\, \left(1-\frac{T_{\rm u}}{\sqrt{T_{\rm u}^2+6.16\times10^{15}\, \rho^{\frac{1}{3}}} } \right)}
  {\left(T_{\rm u}^2+6.16\times10^{15}\, \rho^{\frac{1}{3}}\right)^{\frac{1}{6}}}.
  \label{Ze}
  \end{equation}
The Zel'dovich number as a function of initial temperature is shown in Fig.~\ref{fig_ZevsT}, with different initial densities.
It is seen that it decreases with increasing temperature,
  while the density variation has a minor effect.
Consequently in the following we shall mainly focus on the effect of temperature change on the pulsation behaviors.

\section{Thermal-diffusional pulsation behaviors and flame instability regimes}


\subsection{Numerical simulation setup}

In order to investigate the pulsation behavior of the nuclear flame, we employ direct numerical simulation of the following governing equations
  in the polar coordinate system:
  \begin{align}
  \label{eq_enthalpy0}
  \rho\frac{\partial H}{\partial t} &= \frac{1}{r^2} \frac{\partial}{\partial r} (\kappa r^2\frac{\partial T}{\partial r}) + \rho Q_{\rm c}\Re,\\
  \label{eq_conc0}
   \frac{\partial Y}{\partial t} &=   D\frac{1}{r^2} \frac{\partial}{\partial r} (r^2\frac{\partial Y}{\partial r})-   \Re,
  \end{align}
  with $\kappa$, $D$ and Y being the thermal conductivity, density-weighted mass diffusivity and reactant mass concentration, respectively.
By further adopting the reference frame of the moving spherical flame front, the governing equations are (see Appendix A for details)
  \begin{align}
  \label{eq_enthalpy}
  \frac{\partial H}{\partial \tau} &= \frac{3}{4} \frac{\partial^2 H}{\partial R^2} + \left(1+\frac{3c}{2}\right)\frac{\partial H}{\partial R} +
  Q_{\rm c}\dot{\Re},\\
  \label{eq_conc}
  \frac{\partial Y}{\partial \tau} &= \frac{\partial Y}{\partial R}- \dot{\Re}.
  \end{align}
Here $c\equiv1/R$ (where $R$ is the radius of the spherical flame) is the curvature,
  which is assumed to be constant:
  a planar flame front corresponds to $c=0$,
  the outwardly propagating flames (OPFs) implies positive curvature ($c>0$)
  and it is negative ($c<0$) for inwardly propagating flames (IPFs).
The diffusion term in the concentration equation has been omitted due to the extremely large Lewis number.
As scaling parameters we use the front thickness, its velocity and the temperature of the burnt matter,
  and the EoS (\ref{eos}) reduces to $H=H_0+b T^2$.
The energy release due to the nuclear reaction is
  \begin{equation}
  \label{eq7er}
  \dot{\Re}=\Lambda a^2 \textrm{exp}\left(-\sqrt[6]{\frac{b{E_{\rm a}}^2}{H-H_0}}\right),
  \end{equation}
  where $\Lambda=l_T/u$, with $l_T$ the thickness of the flame thermal diffusion zone and $u$ the stationary flame speed (cf. Appendix A),
  is obtained as a steady state solution to Eqs. (\ref{eq_enthalpy})-(\ref{eq_conc}).
The Dirichlet boundary condition (where $\Delta H = b T^2$), i.e.,
  \begin{eqnarray}
  Y=0|_{R\rightarrow-\infty},Y=1|_{R\rightarrow+\infty},\\
  \Delta H=1|_{R\rightarrow-\infty}, \Delta H={\frac{1}{\theta^2}}|_{R\rightarrow+\infty},
  \end{eqnarray}
  is used to get the steady-state flame as the initial condition for time evolution.

We first find the stationary solution with the front eigenvalue calculated iteratively,
  then solve Eqs. (\ref{eq_enthalpy})-(\ref{eq7er}) with finite difference spatial discretization on uniform/non-uniform grid.
Grid resolution ensures 50-100 numerical points per half width of the energy release,
  and fixed time step using the second-order Runge-Kutta method is adopted.

\subsection{Simulation results for planar flames}

In the WD nuclear flame with $Le=10^7$, the Lewis number can be treated as infinity
  such that the theoretical critical $Ze$ can be used.
While being $Ze_c=4+2\sqrt5$ for ideal gas \citep{siva77,zel85},
  the theoretical critical $Ze$ for WD is twice this value because of the different EoS relevant,
  i.e. \citep{byc95},
\begin{equation}
  Ze_c=2(4+2\sqrt5).
  \label{critical Ze}
\end{equation}
It is also noted that the primary result of this thermal-diffusional pulsation instability in WD is that
  it exists with an average flame speed lower than the laminar flame speed \citep{byc95}.
It is further discussed in \citet{gla13},
  in which they concluded that the carbon flame is pulsatingly stable for the initial condition of WD burning.
\\

\begin{figure}[H]
\centering
  \includegraphics[width=0.85\textwidth]{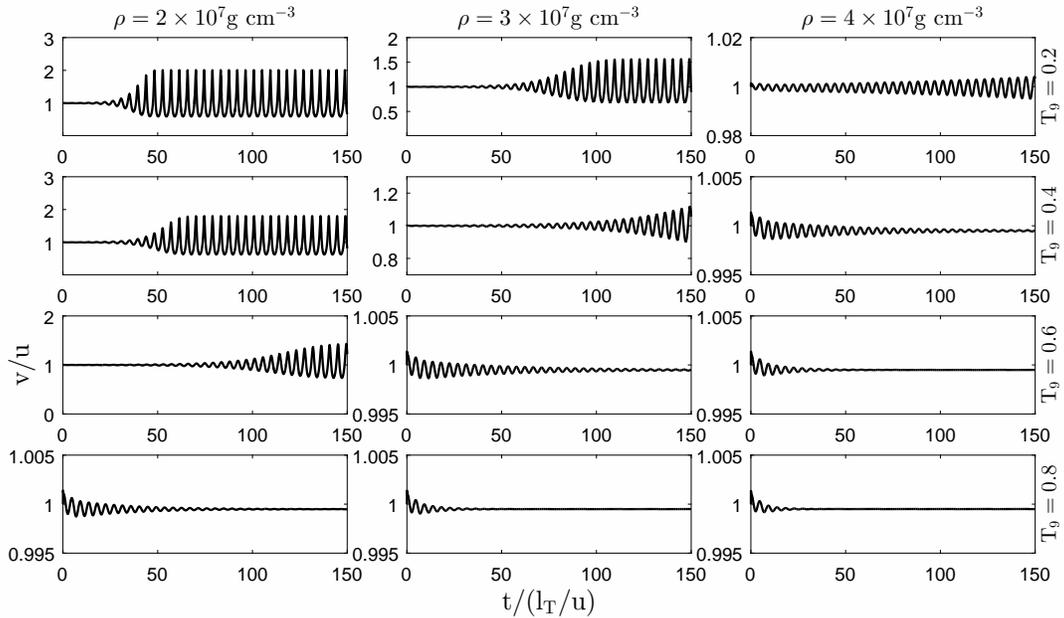}
  \caption{Pulsation behaviors of planar carbon flames with $\rho=2$, $3$, $4\times10^7$ ${\rm g/cm^3}$ and $T_9=0.2$, 0.4, 0.6 and 0.8.
  \label{fig2}}
\end{figure}

The typical density of burning matters is within the range of $2\sim4\times10^7~\rm{g/cm^3}$,
  while the initial temperature of the carbon flame is about $0.6\times10^9$~K \citep[e.g.][]{Woo11};
  so in Fig. \ref{fig2} we show the front velocity evolution for $\rho=2$, $3$, and $4\times 10^7$ ${\rm g/cm^3}$ with various temperatures around $0.6\times10^9$~K.
It is seen from the first column of Fig. \ref{fig2} that for carbon burning at $\rho=2\times10^7$ ${\rm g/cm^3}$ and $T=0.2$ and $0.4\times 10^9$ K,
  pulsation exists with the maximal velocity amplitude being around twice the stationary front speed.
However, the temperature increase leads to a corresponding decrease of $Ze$, so that the flame front becomes more stable:
  for $\rho=2\times10^7$ ${\rm g/cm^3}$, at $T=0.6\times 10^9$ K the flame shows the onset of the unstable regime with gradually growing oscillation amplitudes;
  and at $T=0.8\times 10^9$ K it demonstrates a pulsatingly stable front.
The other two columns of Fig. \ref{fig2} show the flame behavior with higher densities of $\rho=3$ and $4\times10^7 {\rm g/{cm}^3}$.
As the density increases, the numerical critical $Ze$ becomes larger; so for the same temperature, the pulsation of flame becomes weak, or even vanishes.

\begin{figure}[H]
\centering
  \includegraphics[width=0.70\textwidth]{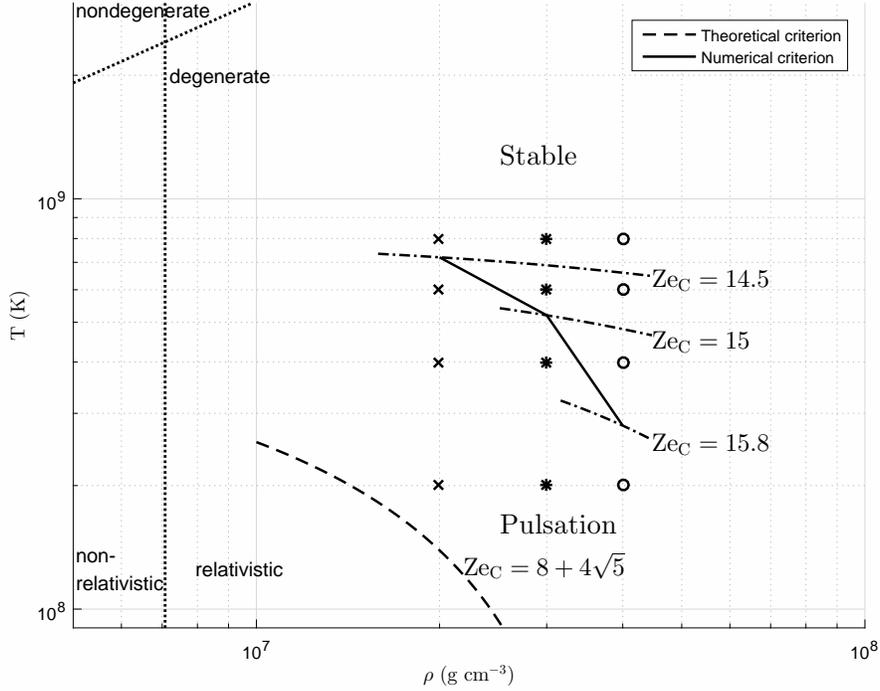}
  \caption{Temperature-density diagram for thermal-diffusional pulsation instability in WD carbon flame.
  The vertical dotted line divides non-relativistic and relativistic regimes due to density,
    while the upper oblique line divides the non-degenerate and degenerate regime \citep{Tim92}.
  Points with their pulsation behaviors shown in Fig. 2 are labeled here with different initial temperatures, i.e., $T=0.2$, $0.4$, $0.6$ and $0.8 \times10^9$ K, and different initial densities, i.e, $\rho=2$, $3$, and $4\times10^7 {\rm g/{cm}^3}$.
  The dashed curve stands for the analytical critical $Ze$ (\ref{critical Ze})
    and the solid line is the critical line of pulsation obtained from simulations at the three initial densities.
  Above the critical line the flame is stable and below it the flame pulsates.
  The dash-dot curves represent the same $Ze$.
  \label{fig1}}
\end{figure}

In Fig. \ref{fig1} we connect the critical points of flame pulsation in different densities with a solid line in the WD temperature-density diagram,
  in order to identify different regimes of pulsation instability in carbon flames.
The typical WD matter condition ($\rho\sim2\times 10^7$ ${\rm g/cm^3}$ and $T\sim0.6\times10^9$ K,), corresponding to relativistic degenerate gas,
  is pulsatingly unstable; and the increase of density and temperature will restrain the pulsation behavior.
The dashed line stands for the analytical $Ze$ (\ref{critical Ze}) based on expression (\ref{Ze}),
  that yields higher value than the one obtained in simulations as a result of the delta-function approximation used in the analysis \citep{Modestov_2011}.
As the initial density increases, the specific heat becomes smaller, see expression (\ref{specific heat}),
  and as such leads to a larger temperature ratio between the burnt and unburnt matter, which is closer to the hypothesis of delta function from analysis.
The growth of the numerical critical $Ze$ (as density increases) conforms this trend.
It is also noted that the density decrease in the outer part of the star may switch the matter into the non-relativistic
  states in which the flame pulsation behavior can vary.
However in this work we are interested in the carbon nuclear burning in the star core region.
\\

\subsection{Numerical simulation of spherical flames}

Naturally the supernova explosion develops in spherical geometry, especially during the initial stage of the flame.
As shown in Appendix B, the corresponding critical $Ze$ \emph{of non-degenerate matter} accounting for the flame front curvature can be given
  theoretically.
According to the expression (\ref{conclusion_Ma_OPF}) and reasonable curvature values of $\sim 0.05$,
  the critical $Ze$ (stability limit) can be reduced by 10-20\% in OPF.
If this trend keeps in degenerate WD matters, it becomes crucial in the kernel ignition stage when the magnitude of the curvature is large,
  as the highly unstable pulsating flame can be quenched \citep[][see also the Appendix B]{sun02,Law03}.
Consequently it implies that the flammability limit for flames with a certain Zel'dovich number:
  for flame kernels with its radius smaller than a critical value for which the curved flame quenches due to pulsation,
  carbon flames cannot be ignited.
However after ignition,
  as the nuclear flame propagates the curvature effect vanishes with the growth of the flame radius.



As the front curvature plays an important role in the pulsating instability of the WD flame,
  we take it into consideration in the following simulations.
Figure 4 shows results for carbon flames propagating with different curvatures,
  with variable temperatures around $T=0.6\times10^9$ K and at the fixed density $\rho=2\times 10^7$ ${\rm g/cm^3}$.
From the right panels it is seen that for $T=0.8\times10^9$ K,
  the flame is stable in IPF and planar geometry, becomes unstable in OPF with curvature $c=0.01$,
  and pulsates more dramatically at $c=0.05$.
Similar trend can be found for $T=0.6\times10^9$ K as seen from the centra column.
For $T=0.2\times10^9$ K, it is shown that the pulsating carbon flame in planar geometry
  will quench due to dramatic pulsations in the OPF with $c=0.05$.
This quenching behavior demonstrates that for
  $\rho=2\times 10^7$ ${\rm g/cm^3}$ and $T=0.2\times10^9$ K carbon flame,
  the local hot kernel should have a radius much larger than $\sim20$ cm (correspondingly $c<0.05$)
  so that it does not quench as a result of the dramatic curvature induced pulsation.
Similar to planar flames, for flames with the same curvature, pulsation amplitude grows with decreasing environment temperature.

\begin{figure}[H]
\centering
  \includegraphics[width=0.85\textwidth]{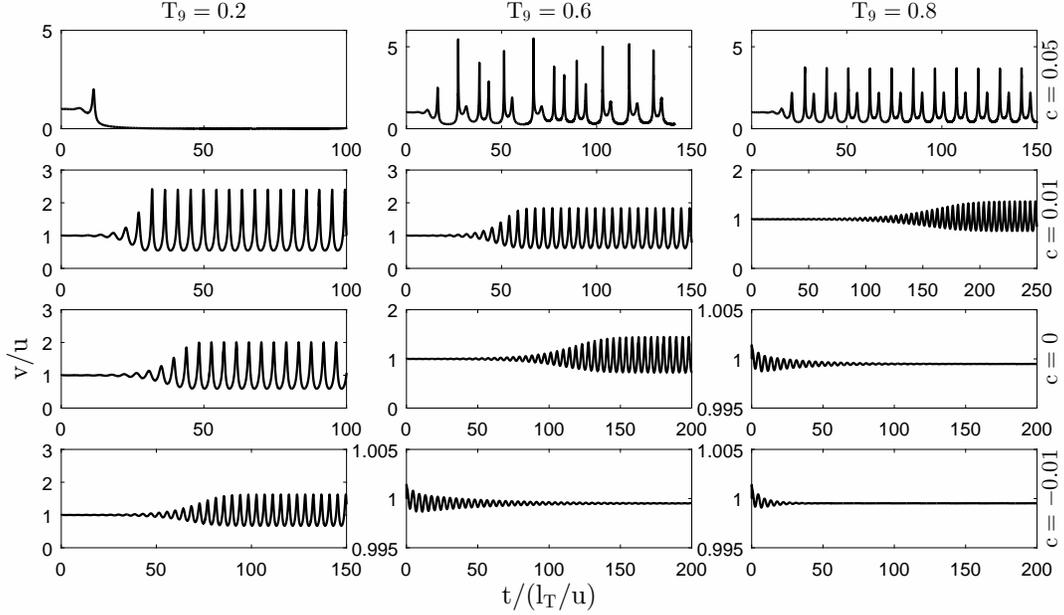}
  \caption{Pulsation behaviors for spherical carbon flames with $\rho=2\times 10^7$ ${\rm g/cm^3}$ and various curvature values at different temperatures. }
  \label{fig001}
\end{figure}

Also at density $\rho = 2 \times 10^7$ ${\rm g/{cm}^3}$, through numerical simulations,
  we found the critical Zel'dovich numbers of pulsation for spherical carbon flame at different curvatures (Fig. 5).
It is seen that for OPF, the critical Zel'dovich number ($Ze_C$) decreases as the curvature $c$ increases, making it easier to pulsate;
  a quenching critical Zel'dovich number ($Ze'_C$) is found for the large curvature flame of $c=0.05$.
For flames with Zel'dovich number larger than $Ze'_C$, i.e. below the point of $Ze'_C$ in Fig. 5, flame quenches due to pulsation.
As the Zel'dovich number has a negative dependence on the initial temperature (see Fig. 1), at a certain curvature,
  the decrease of temperature will facilitate flame pulsation, similar to the planar flames (Fig. 3).

\begin{figure}[H]
\centering
  \includegraphics[width=0.75\textwidth]{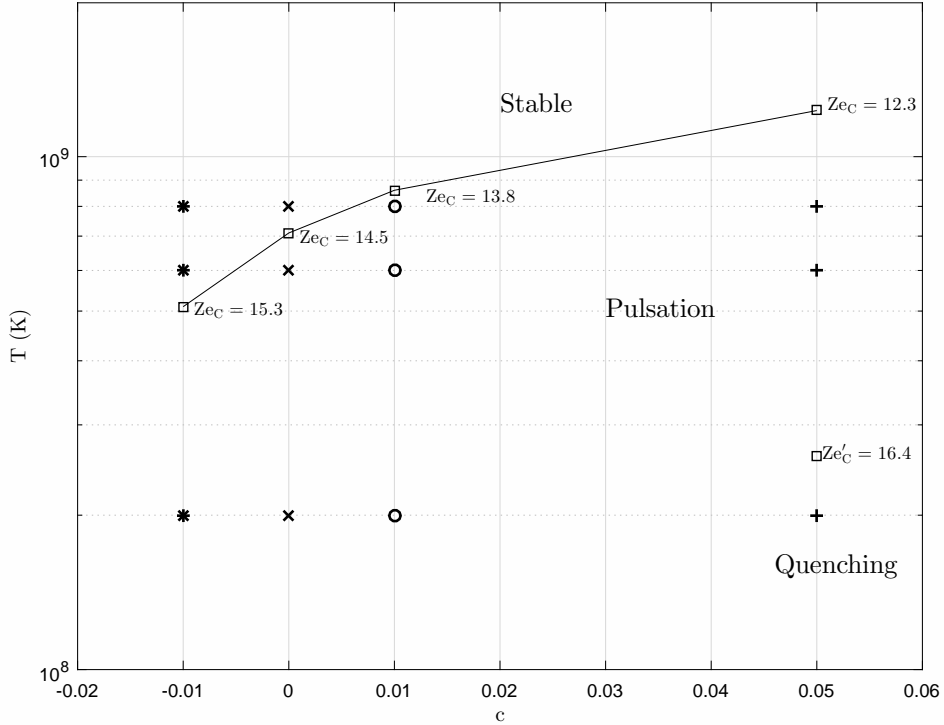}
  \caption{Pulsation regimes for flames with different curvatures at the specific density $\rho = 2 \times 10^7$ ${\rm g/{cm}^3}$.
The solid line connects the critical temperatures of pulsation (with corresponding critical Zel'dovich number $Ze_C$ also labeled),
  above which the flame is stable and below it the flame pulsates.
The square marker with Zel'dovich number $Ze'_C=16.4$ represents the critical temperature of flame quenching at $c=0.05$,
  below which the flame quenches.
  \label{figc}}
\end{figure}

\subsection{Summary}

According to the above analysis and simulations, we can draw the following conclusions for the thermal-diffusional pulsations in WD flames:
(1) The critical Zel'dovich number of planar carbon flames with density $\rho=2\times 10^7$ ${\rm g/cm^3}$ is $14.5$,
  about $15\%$ lower than the theoretical value derived based on the delta function reaction rate.
As the density increases, the critical Zel'dovich number increases and the pulsation instability is restrained.
(2) Carbon flames with density $\rho=2\times 10^7$ ${\rm g/cm^3}$ is pulsatingly unstable at temperature $T=0.6\times10^9$ K;
  while flames with higher temperatures (e.g., $T=0.8\times10^9$ K) tend to be stable.
(3) Both lower temperature and positive curvature (OPF) can enlarge the amplitude of pulsation.
For example, flames with $\rho=2\times 10^7$ ${\rm g/cm^3}$, $T=0.6\times10^9$ K and $c=0.05$
  pulsate with the pulsation amplitude as large as more than six times the stationary propagation speed.
(4) At the density $\rho=2\times 10^7$ ${\rm g/cm^3}$, the normally pulsating carbon flame with \textbf{$T=0.2\times10^9$} K
  can be quenched in OPF if the hot flame kernel is not large enough;
  leading to a critical radius of $\sim20$ cm under which the flame cannot be ignited.

\section{Discussions}

For spherical flames, we have investigated the stability of a flame which propagates with its velocity at a fixed given radius (curvature).
This assumption is justified by relatively small curvature values,
  implying that the flame radius is much larger than the front thickness and the flame does not move far from this fixed radius during the pulsations.
This approximation is valid even for the largest curvature ($c=0.05$) used in the simulations,
  in which the distance of flame propagation before the pulsation is not far ( $\sim$ 10 times the flame thickness as shown in Fig. 4 upper panels)
  from the fixed flame radius.

Furthermore, the analytical form of EoS, with assumptions that the WD core region has slightly variations from
  the ultra-relativistic and fully degenerate EoS, has been used here;
  this is convenient for comparisons with previous results but may lead to differences in the results compared with the real situation.
A more realistic EoS \citep[e.g.,][]{timmes1999} should be used in further numerical simulations.
It should also be noted that the variation of the EoS during flame propagation is not considered.
Generally speaking, as the flame propagates into the non-degenerate part of the WD where the critical $Ze$ value is 1/2 smaller,
  it induces more violent pulsations if the flame is not quenched during the transition of EoS.

Additionally, the assumption of WD being composed of pure carbon is made;
  and only the pulsation of the carbon flame is considered here, without considering the oxygen flame.
This is usually reasonable as the thickness of the oxygen flame is much larger and the effective Zel'dovich number is small,
  i.e., pulsation of oxygen flame itself is less violent than carbon flame.
However, if the detailed reaction mechanism of carbon is considered,
  and further reactions involving oxygen and beyond is included,
  the pulsation behaviors of the WD flame may vary from the results shown here.

Further studies would also involve the 2D effect of the pulsation instability in which the transverse waves may be observed;
  and the compressible effect of pulsation, leading to the formation of pressure waves and shock waves.


\acknowledgments
We acknowledge the anonymous referee for his/her constructive suggestions that help in improving the paper.
YG thanks Prof. Zheng Chen from Beijing University for discussions on spherical flame propagation.
This work was supported by the National Science Foundation of China grant 51206088.
YG also acknowledges support from the Tsinghua-Santander Program for young faculty performing research abroad.
MM acknowledges the Swedish Research Council for the International Postdoc grant No. 637-2014-465.

\appendix

\section{Appendix: Nondimensionalization of governing equations  } \addvspace{10pt}
In order to simplify the problem, we introduce the non-dimensional forms of governing equations (10) and (11).
The EoS (\ref{eos}) can be simplified to $H=H_0+b T^2$, with $b$ denoting the coefficient of the temperature term;
  the thermal conductivity $\kappa$ has the form \citep{yak80,byc95}
  \begin{equation}
  \label{thermal conductivity}
  \kappa=\frac{0.81\hbar c_{\rm l}^2 k^2}{e^4Z\Lambda_{ei}}{(\rho N)}^{1/3}T=\kappa_0 T,
  \end{equation}
  with $\Lambda_{ei}$ being the Coulomb logarithm; and the specific heat (\ref{specific heat}) is $C_p=C_0 T$.
The total energy (\ref{total energy}) can then be written as
  \begin{equation}
  \label{}
  H=H_0+b T^2=\frac{4}{3}E=\frac{4}{3}E_0+\frac{2}{3}C_0T^2,
  \end{equation}
  and the thermal diffusion term in equation (10) can be reduced to
  \begin{equation}
   \frac{1}{r^2}\frac{\partial}{\partial r}(\kappa r^2\frac{\partial T}{\partial r})=\kappa_0(\frac{1}{r}\frac{\partial T^2}{\partial r}+ \frac{1}{2} \frac{\partial^2 T^2}{\partial r^2}).
  \end{equation}
Then by introducing the enthalpy $H$, equation (10) is reduced to
  \begin{equation}
  \label{eq_enthalpy0_A2}
  \rho\frac{\partial H}{\partial t'} =\rho u\frac{\partial H}{\partial r} + \frac{\kappa_0}{b}(\frac{1}{r}\frac{\partial H}{\partial r}+ \frac{1}{2} \frac{\partial^2 H}{\partial r^2}) + \rho Q_{\rm c}\Re.
  \end{equation}
Noting that here the reference coordinate system has been changed to be co-moving with the flame front ($t \rightarrow t'$),
  so that there is one additional convection term with $u$ representing the flame speed:
  \begin{equation}
   \frac{\partial H}{\partial t}=\frac{\partial H}{\partial t'}-u\frac{\partial H}{\partial r}.
  \end{equation}
In the dimensionless framework defined as: $R=r/l_T$ and $\tau=t'/(l_T/u)$,
  with the length scale $l_T$ being the thickness of the thermal diffusion zone $l_T=\kappa/(\rho u C_p)$,
  the governing equations (10, 11) are finally converted to
  \begin{align}
  \frac{\partial H}{\partial \tau} &= \frac{3}{4} \frac{\partial^2 H}{\partial R^2} + \left(1+\frac{3c}{2}\right)\frac{\partial H}{\partial R} + Q_{\rm c}\dot{\Re},\\
  \frac{\partial Y}{\partial \tau} &= \frac{\partial Y}{\partial R}- \dot{\Re},
  \end{align}
  where $\dot{\Re}=(l_T/u)\Re$ and $c=1/R$ is the flame front curvature.

\section{Appendix: Curvature effect of flame thermal-diffusional instability for a non-degenerate ideal gas} \addvspace{10pt}


The influence of stretch on the pulsating instability of $Le\gg1$ flames in ideal gas,
  identified by the critical Zel'dovich numbers, is specified in the following.

In the reactive fluid system considered,
  the chemical/nuclear reaction is simplified as a one-step reaction described by the Arrhenius' reaction rate equation
  $W(Y,T)=z\rho Y {\rm exp}({-\frac{E_{\rm a}'}{R^{0}T}})$,
  where $z$ is a frequency factor, $Y$ the concentration of the controlling reactant, $E_{\rm a}'$ the activation energy and $R^{0}$ the universal gas constant.
Accordingly and following the definition (6), the Zel'dovich number is noted as:
  \begin{eqnarray}
  \nonumber
  Ze=\frac{E_{\rm a}'}{R^{0}T_{\rm ad}}(1-\frac{T_{\rm 0}}{T_{\rm ad}}),
  \end{eqnarray}
  with $T_{\rm 0}$ being the initial temperature and $T_{\rm ad}$ the adiabatic flame temperature.
The Arrhenius form of the reaction rate is adopted for a more general application of the conclusion made here,
  while the nuclear flame discussed in the main text applies to the following conclusions in the context of the same definition of Zel'dovich number.

The temperature and concentration distributions in the one-dimensional spherical flame with unburnt medium at rest yield the following heat
  and mass diffusion equations \citep{siva77}:
  \begin{eqnarray}
  \nonumber
  \rho C_{\rm p}\frac{\partial T}{\partial t}= \frac{1}{r^{2}} \frac{\partial}{\partial r} (\kappa r^{2}\frac{\partial T}{\partial r})+qW(Y,T),\\
  \rho \frac{\partial Y}{\partial t}=\rho D \frac{1}{r^{2}} \frac{\partial}{\partial r} (r^{2} \frac{\partial Y}{\partial r})-W(Y,T).
  \label{Gov}
  \end{eqnarray}
Here the thermal conductivity $\kappa$, density-weighted mass diffusivity $D$ and specific heat $C_{\rm p}$ are assumed to be constant.
The boundary conditions for this problem are:
  \begin{eqnarray}
  \nonumber
  T = T_{\rm b}, \quad Y=0, \quad &(r\leq R),\\
  T\rightarrow T_{\rm 0}, \quad Y\rightarrow Y_{\rm 0}, \quad &(r\rightarrow+\infty),
  \end{eqnarray}
  where $T_{\rm b}$ is the burnt flame temperature, $Y_{\rm 0}$ the initial reactant concentration
  and $R(t)$ the time-dependent flame front position.

Under the assumption that the flame radius is much larger than the flame thickness as well as the flame perturbation,
  and by adopting proper boundary conditions in the burnt and unburnt regions and jump conditions across the flame front,
  stationary solutions can be readily derived.
Then the small harmonic perturbation analysis is carried out,
  showing that the criterion for pulsating instability in OPF with $Le\gg 1$ is
  \begin{equation}
  Ze\geq (4+2\sqrt{5})[1-(1+\frac{2}{\sqrt{5}}-2\epsilon)\frac{\kappa'}{\dot{R}^2}]\label{conclusion_kappa_OPF},
  \end{equation}
  where $\kappa'=2c\cdot\dot{R}$ is the stretch rate, and $\epsilon=T_{\rm 0}/T_{\rm ad}$ is the temperature ratio.
This criterion degenerates to the result in planar flames where $\kappa'=0$ \citep{siva77}.
It is clearly seen that for the OPF, the flame is less stable as $\epsilon$ is usually of the order $\sim 0.1$.
The stretch effect is coupled to the flame speed \citep[][]{matalon2003}, i.e.,
  $\dot{R}=\sigma(1-2\sigma l^{\rm b}/R)=\sigma(1-2\sigma \rm{Ma}\cdot c)$,
  where the Markstein length $l^{\rm b}$ represents the ratio between the change in the laminar burning speed and the flame stretch rate,
  and the Markstein number Ma takes values around unity and can be treated as a constant for the specific reaction and physical conditions considered \citep{matalon2003,Dur03}.
Here $\sigma=T_{\rm b}/T_{\rm ad}$ is the thermal expansion parameter, which deviates from unity in $Le\neq1$ flames as a result of the flame curvature
  \citep{che10}.
Then the instability criterion for OPF ($c>0$) and IPF ($c<0$) is further written as:
  \begin{equation}
  Ze\geq(4+2\sqrt{5})[1-(2+\frac{4}{\sqrt{5}}-4\epsilon)c]\label{conclusion_Ma_OPF}.
  \end{equation}

\begin{figure}[H]
	\centering
	\includegraphics[width=0.5\textwidth]{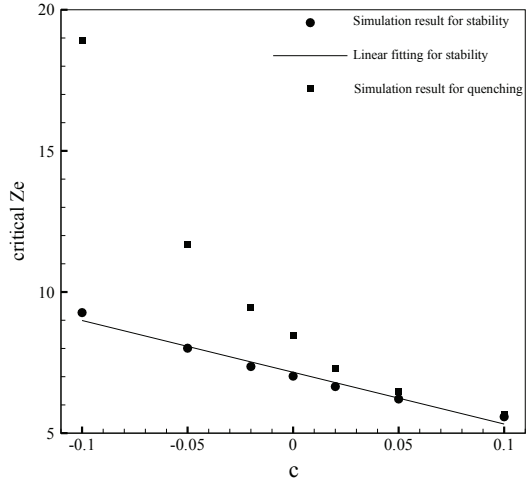}
	\caption{Thermal-diffusional pulsation regimes for ideal gas.
  Critical $Ze$ for pulsation (circles) and quenching (squares) as a function of curvature.
  Below the circles the flame is stable; in between the circles and squares the flame pulsates; above the squares the flame quenches due to severe pulsations.}
	\label{cZe}
\end{figure}

Direct numerical simulations based on the governing equations (\ref{Gov}) is also adopted for a fixed curvature $c$.
The simulation and analytical results of critical $Ze$ for different curvatures are shown as circular dots and the solid line
  respectively in Fig. ~\ref{cZe}.
The critical $Ze$ obtained from the simulations reduces approximately linearly from $\sim9$ to $\sim5$ as the curvature increases from IPF with $c=-0.1$
  to OPF with $c=0.1$,
  meaning that negative curvature retards the instability and positive curvature facilitates it.
The critical $Ze$ derived from asymptotic analysis (\ref{conclusion_Ma_OPF}) is shown as a reference line in the figure,
  and it is noted that analysis always yields larger critical values.
For a fixed curvature, flame extinction occurs as a result of severe pulsation if the reaction is highly temperature sensitive,
  i.e., $Ze$ is too large.
We explore this possibility and plot the squares in Fig. ~\ref{cZe} representing the quenching limit of $Ze$,
  beyond which the flame quenches.
Thus the $Ze-c$ plane is separated into three regions:
  below the filled circles the flame is stable,
  between the circles and squares the flame pulsates,
  and above the squares the flame quenches.
Flame with constant Zel'dovich number (e.g. $Ze=7$) transitions from being stable to unstable until quenching,
  purely due to the change of flame curvature, i.e. from IPF to planar to OPF respectively.

\setcounter{figure}{0}
\renewcommand{\thefigure}{A\arabic{figure}}


\clearpage

\end{document}